# A normative account of defeasible and probabilistic inference

Julio Lemos
Max Planck Institute - Hamburg / University of São Paulo
lazzarini.lemos@gmail.com

## 1. Introduction

A good explanation -- among many others -- of logical consequence is that given a possibly empty set of assumptions *A* that are said to entail *x*, written $A \vDash x$, one is *bound* to believe *x* whenever it is the case that she believes the conjunction of all elements of *A*. This explanation shows a weak correspondence with the Tarskian semantic definition of logical consequence, according to which *A* entails *x* if and only if *x* is true under all interpretations (models) that make all elements of *A* true. But the relevant term here is "bound" ("is obliged to"), and the talk is of a normative conception of logical consequence.

But where does this command come from? To understand some of the implications of what we call here a normative account of logical consequence, we will consider a notion which is not very well understood until the present day: that of "defeasible subsumption". To grasp this notion, it suffices to consider the following example. Suppose a creature is found which appears in almost every way to be some kind of alligator. For immediately the founder notes that it is strangely reddish -- neither brown nor green. Can it be considered to be an alligator? Well: it seems that unless proven otherwise we *are allowed to conclude* that it is a member of the said species. Perhaps it is only some weird instance thereof. But notice that now we most reasonably used "allowed" (to conclude) instead of "bound".

Let us draw here an easy analogy. In mathematics, once defined, a notion is to be unequivocally applied to all the cases it defines (the *definienda*). For instance a set *A* is defined to be modular if and only if (iff) for any members *x*, *y*, it is the case that $x < y$ iff the rank of *x* is lower than the rank of *y* (according to a function that assigns to every element of A an element of the set of real numbers), written $r(x) < r(y)$. Under the right kind of assumptions, the set of modular sets X subsumes any modular set Y as long as the above condition is satisfied. One is not only allowed to conclude that a given set A is modular, i. e., that A is subsumed by X (in our parlance), but most accurately *bound* to reach such conclusion. It is not so in the case of the natural sciences; and most naturally not so in the case of taxonomy. Even the most strict code of scientific practices will not go as far as to force a taxonomist to classify a red alligator as a regular alligator (supposing our example is not entirely flawed). He may be allowed to go on to say "well, according to my expertise that is an alligator, albeit a strange one", but he is not bound by the classification. Non-mathematical taxonomies are very important scientific instruments since the times of Aristotle, but they are always more or less flawed. So is common sense. But we suspect that side by side there are cases in which there is a clear 'obligation to infer', event outside mathematical boundaries (practical inferences, such as those mandated by law or moral principles).

## 2. A normative conception of logical consequence

It appears that John MacFarlane is the first logician to undertake the risk of a detailed analysis of the intuition according to which logic has a normative character, constraining reasoning. His claim is that it is possible to somewhat forget this appeal to intuition and really grasp how logic provides norms for reasoning, that is, in what sense logic is normative for thought. He proposes two "bridge principles", a bridge being a consequence relation between a "logical fact" and an explaining claim in deontic terms: if $A, B \vDash C$, then it follows that my belief concerning A, B and C is bound or permitted (conditionally) to be such and such.

His first bridge principle results from analysing and selecting the best permutations on a table, after having scrutinized them one by one: *if $A, B \vDash C$, then (normative claim about believing A, B, and C)*. This normative claim could be that (Co+) if you believe A and you believe B, you ought to believe C; (Co-) if you believe A and you believe B, you ought not disbelieve C; and so on. The permutations happen over three different possibilities (deontic operator embedded in consequent (C), in both antecedent and consequent (B) or scoping over the whole conditional (W)), each of which subdivide into three possibilities (deontic operator is a strict obligation (o), a permission (p) or a defeasible reason (r)), each of which finally subdivide into two possibilities (polarity of 'to believe': either to believe (+) or not to disbelieve (-)). It may sound a bit confusing, but MacFarlane's strategy was utter *divide et impera*. He then consider four criteria for rejecting or accepting each version of the normativity claim: the problem of excessive demands, the paradox of the preface, the strictness test, the priority question and logical obtuseness, for which we do not have time here. His preferred and well argued versions are Wo- ("you ought to see to it that if you believe *A* and you believe *B*, you do not disbelieve *C*) and Wr+ ("you have reason to see to it that if you believe *A* and you believe *B*, you believe *C*). For now we will take his guesses on faith, specially because it is not a controversial claim or result, but only a methodology we chose to stick to.

The second of his bridge schemata is more interesting for our purposes, and concerns formality: *If the schema S is formally valid and you apprehend the inference A, B / C as an instance of S, then (normative claim about believing A, B, and C)*. Formality is perhaps the most important feature of logic; moreover, the relationship between formality and defeasible logic is still mysterious, since the very reason for defeasibility to be called upon is that it often deals with empirical notions such as "information available to agent x at the time" and "what has been supported by evidence or expert opinion" -- notions that are *unheimlich* (something only a German adjective may sometimes convey) to each other. MacFarlane's hypothesis in [Mac2004], which notwithstanding does not appear to be fully endorsed, is that "we require logical validity to be formal because we require it to be transparent, and we require it to be transparent because of the reasons and responsibilities to which it gives rise". The point is that grasping an inference schema is the same as grasping a formal structure; therefore transparency. And if one *sees* it clearly that some reasoning is an instance of an inference schema, one has no rational justification for abstaining from inferring. Compare the following inferences:

> (1) Cicero talked the talk and Tully walked the walk; therefore, someone walked the walk and talked the talk.

> (2) Cicero talked the talk and Cicero walked the walk; therefore, someone walked the walk and talked the talk.

The difference between them is that, in (1) it is not clear that the consequent follows from the antecedent *unless one knows that Cicero is just another name for Tully*. It is therefore not transparent; and that is the reason why philosophers and logicians often claim (1) not to be an instance of a valid inference, to wit: *Fa, Ga / ∃x (Fx ∧ Gx)*. On the other hand, (2) does not suffer from this problem. Substituting the name/object Cicero in the domain of interpretation for the constant *a* and each predicate for a different symbol among *G*, *F* in the scheme above, we obtain a valid "concrete" inference, so that one does need to know what a "Cicero", walking, talking is. One only needs to recognize the basic logical vocabulary ("something is a ... and something is a …, therefore at least one thing is a ... and is a …").This transparency of form is what accounts for the obligation to believe the consequent, as long as one believes the antecedent, given that consequent and antecedent happen to instantiate an inference class. "An obvious entailment is one that an agent *ought* to see", summarized Hartry Field in [Ha2009]. So much for MacFarlane's account.

That is what in turn allows what in Computer Science is called *machine readability*. A computer program -- sometimes rightly called a 'logical agent' -- can only reason if given an entirely transparent instance of an inference scheme. Otherwise it will report an error. Transparency then sets a lower bound on inference capability, and therefore creates, so to say, an obligation to infer (irregardless of computational complexity, even a machine is able to infer given this degree of obviousness!). Logic, being itself an agent, operates automatically, inferring each and every statement it is allowed to infer, as long as there is a (conditional or otherwise) command or trigger.[1] This allowance, in human terms, is seen as an obligation, as long as transparency is given. What does not count as transparent is defined by way of

---

[1] To specify further conditions, and even to try to define precisely what is allowed here by the logic, would be to take sides in some kind of Platonism/Anti-platonism dispute. What we have in mind is a simple feature of Logic: a logical agent virtually infers everything allowed by the underlying logic, but there is no tangible, or maybe not even intangible, world where these inferences take place. Given *A*, the logic would allow itself to infer *A*, *A* ∧ *A*, *A* ∧ *A* ∧ *A* , *A* ∨*B*, *A* ∨ *C*, and so on, until and after the world comes to an end. A Platonist would note that these inferences are there somewhere, *done*, but one doesn't have access to all of them in time (or at the same time), and so forth. This is not a problem at all, specially for the Anti-platonist. The inferences out there, potentially infinite, as follows from the said constrained 'allowance', have a practical use: when one needs them, building a computer program or reasoning by himself, on one's daily life or writing a paper on Logic, it suffices to trigger and select them. Therefore the talk of a command or trigger; the fact of effective inferences taking place, immediately requires one to deal with resource limitations.

This problem is related to a common observation, to wit: we are able to define with precision the operations of a Turing machine, but we will have a lot of trouble trying to define what is, abstractedly, an algorithm. Notwithstanding, according to Church's Thesis, there is a correspondence between the usual informal notion of an algorithm and the standard formal characterizations of it by Kleene, Turing and others. See [Rogers1967] and [Post1944].

complementariness by human limitations: given, for instance, a thousand premises, it is easy to see that one is not obligated (although trivially permitted to) to *see* all that comes out of the box. Similarly, computers are limited by available computational resources: whenever trivial inferences are allowed and triggered, their number is virtually infinite. The so called resource-bounded reasoning tasks were idealized to account for, and deal with, human and computer limitations.[2]

Seeing logic as an agent actually helps to understand much of what goes on in the philosophy of logic. The constraints acting upon an agent in a belief revision situation (which is what logic is for) are the exact constraints that the underlying logic imposes on itself. A classical logic agent facing (2) is not authorized to infer by the scheme *Fa, Ga / ∃x (Fx ∧ Gx)* because it is not authorized -- as long as it does not happen to be out of order -- to see (2) as an instance of *Fa, Ga / ∃x (Fx ∧ Gx)*. The prohibition relies on the difference between Cicero and Tully; they *might* represent the same object in the domain of interpretation, but they are not *shown* to be the same (it is not transparent that they are the same). If the agent reportedly received the information that Cicero is Tully and believes it, he may add this premise to his reasoning and consequently add *a = b* to his candidate derived inference scheme *Fa, Ga, a = b / ∃x (Fx ∧ Gx)* whenever his underlying logic is classical logic with equality; and then he contemplates the possibility that the statement

> (3) Cicero talked the talk and Tully walked the walk and Cicero is Tully; therefore, someone walked the walk and talked the talk may be an instance of the enriched valid scheme *Fa, Ga, a = b / ∃x (Fx ∧ Gx)*. From the extended bridge principle: If Schema *S* is valid and you apprehend the inference *A, B, C / D* as an instance of *S*, then you ought to see to it that if you believe *A* and you believe *B* and you believe *C*, you believe *D*, the agent is now not only contemplating the possibility of (3) but, having apprehended (3) to be an instance of a valid scheme, *obliged* to arrive at the consequent in (3), something which was impossible given the faulty premises in (2), which has the same consequent.

A machine would ask the agent to formalize (3), and then it would match (using a proper matching algorithm) the formalization of (3) to a scheme *S* stored somewhere in its memory. Then the machine would automatically recognize the valid inference, allowing the conclusion that *D* "is the case" without knowing what in the world "to be the case" *means*. A human agent would then translate *D* to "someone walked the walk and talked the talk", seeing that it is the same conclusion: it is the case that someone walked the walk and talked the talk. Deterministic operation (one may think of a theorem prover) is a good machine illustration of philosophical necessity and of moral obligation -- no wonder operators like deontic *O* and modal *N* share an analogous role in deontic and modal logic, although both reveal to be very troublesome in real life cases.

---
[2] See note above.

## 3. Defeasible and probabilistic inferences. Conclusion

Are we sometimes *obliged* to infer non-monotonically, that is, to abandon a conclusion in the presence of new information, or either prematurely drawing conclusions alleging we do not have time to wait for new information?

The classical example of the literature is about swans. Someone who is an expert in swans somewhere in Europe may have grown to firmly believe that all swans are white. He believes that φ: if *x* is a swan, then *x* is white. Suppose the expert_1 (let us call her so) is a classical reasoner. The appearance of a black swan will force the expert to immediately change his belief set, dropping φ (and consequently all the explicit consequences of φ being true). Suppose now the expert has a friend, expert_2, who reasons non-monotonically. Expert_2, facing the mysterious appearance of the black swan (he has never seen one), will reason differently: if *x* is a swan, then one is allowed to infer that *x* is white, unless the statistics on the population of swans is shown to have changed. Begging to disagree with his friend, expert_2 does not drop φ. Who is right? What is preventing expert_1 from keeping φ? What is allowing expert_2 to keep his belief in φ? There is still another alternative for the non-monotonical reasoner: to check first whether *x* is not non-white and, if *x* happens to fulfill this condition, infer that *x* is white. That would trivialize the inference: white(swan) ∧ ¬(¬(white(swan)) ⊨ white(swan) ≡ (*p* ∧ ¬¬*p* ⊨ *p*) ≡ (*p* ∧ *p* ⊨ *p*) is a tautology, a sentence which is entirely independent of φ.

A basis for expert_2's inferences might be the first form of Bayes' Theorem, for instance:

(Bayes' Theorem). $P(x|A) = (P(A|x) * P(x))/P(A)$.

He has a limitation, to wit, that given a possible inference to the hypothesis white(s) = x from the proven antecedent (evidence) swan(s) = A, $P(x|A) = (P(A|x) * P(x))/P(A)$, which reads "the probability of x given A equals ((the probability of x given A times the probability of x) divided by the probability of A)". Take $P(x) = 0.9$ and $P(A) = 1$, assuming that the mysterious swan were captured. Then $P(x|A) = (1 * 0.9)/1 = 0.9$. The rational principle for inferences is not clear. The logic only tells expert_2 that he may keep ϕ, as long as he will check if statistics are unchanged; and if he is not able to check the swans' color, then he is allowed to infer that x. The rational principle applies to the statistics (the domain of discourse and the agent's knowledge reflecting it); he could use a very conservative approach, indeed, inferring x only when the conditional probability is infinitesimally close to 1, as discussed in [Leh1992]. When forced to make a choice in a pressing situation, as in the game below, he will drop this conservative principle and reason more lightly. (Note that 0.9 is not infinitesimally close to 1.) The automatization of this class of probabilistic consequence relations is not without difficulties.

Resource boundedness may thus be a factor to consider when dealing with this very simple problem. Suppose neither expert_1 nor expert_2 has the time to check on the individual swan's color. Forced to make an inference, expert_1 will abstain from inferring that the swan is white. She is in fact *obliged* not to infer, given her logic (remember that expert_1 is a classical agent). On the other hand, expert_2 may or may not infer that the swan is white; given her opinion on the unsurpassable importance of statistics, we may be sure that she will infer that the swan is white. We might even be inclined to state that she is *bound* to do so -- were she a machine.

Let us test the implications of the underlying logic, supposing that guessing the color of swans were a win-lose-draw game. Fancy an expedition has captured 100 swans and held them captive in a dark chamber, and both experts, which we now assume to have the same knowledge, were asked what color are them, based on their reason and knowledge. Expert_2's answer would be on the lines of "white: 99" or maybe "white: 100"; expert_1's would most surely be "white: zero, black: zero" or "null" (that depends on what our interpretation of "not inferring A" is). Assuming the frequency of black swans tends to zero and that there are no other-colored swans (and that is what the classical example assumes), we know that expert_2 will win. Facing resource limitations, non-monotonic logic wins by default. Notice that they share the same knowledge! It is their logics alone that define the outcome of the competition. Or is it? We may say: given knowledge-invariance, the outcome of a knowledge-game depends on the logic alone. But that would be a very different scenario should we introduce a new, and realistic, factor: domain-variance. If both experts were put in an alternate world where the white/black swan frequency were shifted from 99-1 to 1-99 without their knowing so, and the game adhered to a common rule (errors eliminating right answers), expert_2 would probably lose. "Logical luck", someone would say. Correcting our hypothesis, we should say: given knowledge-invariance, the outcome of a knowledge-game is logic and domain-dependent. The underlying domain must correspond to the knowledge factor.

Assuming we could make expert_2's reasoning task entirely transparent, as defined in the previous section, then the answer to our question above is: assuming that one is a non-monotonic agent, she is obliged to infer non-monotonically. That means the agent has no choice: if there is an exception to the rule, then he infers accordingly; if no exception is present, or he is operating on a bounded basis (time pressure, for instance, or simply a situation analogous to "swan-occultation"), then he must infer from the hypothesis. A clear, but negative-polarity definition should be on the lines of:

> **(NonMR-)** If it is transparent that $A$ defeasibly entails $x$, i. e. $K \vDash A \mathrel|\!\sim x$ for a given underlying knowledge base $K$, and there is no information available indicating that $\neg x$, then one is prohibited to believe $A$ without believing $x$.

A positive version:

> **(NonMR+)** If it is transparent that $A$ defeasibly entails $x$, i. e. $K \vDash A \mathrel|\!\sim x$ for an underlying knowledge base K, and there is no information available indicating that $\neg x$, then one is obliged to believe $x$ if one believes $A$.

But how is it that $K \vDash A \mathrel|\!\sim x$ (e. g. instantiated by the inference "given that $s$ is a swan, $s$ is white, unless there is sufficient evidence on the contrary") can be transparent? Earlier we endorsed transparency as formality -- but is defeasibility formal, being grounded on empirical notions such as the frequency of events on a given domain, or the degree of evidence required to establish a statement as acceptable? How can formality be domain-dependent in this complex setting? The question can be addressed by looking at rules outside *K*, but which must be taken into consideration by the reasoning agent. *K* may be seen as a formal ontology, where every element, property and relation in the domain is adequately formalized (there is a map from each symbol in the ontology to an object in the domain), and *R* as a system of rules saying when the agent is allowed, prohibited or obliged to infer. For instance: "in case there is expert testimony in favor of *x* and expert is labeled reliable, then infer *x*". More formal examples would be something like:

> (Example rule). If *X* is an instance of schema $\Gamma \mid \sim \varphi$, then infer the consequent of *X* whenever the antecedent of *X* is accepted as a proven hypothesis and the negation of the consequent of *X* is not proven.

> (Example rule 2). When the conditional probability of $\varphi$ given $\Gamma$ is infinitesimally close to 1, infer $\varphi$.

Logic is then seen as the invariant part of an agent, commanding inferences from a knowledge base according to a set of rules. Both *K* and *R* change, but the agent's logic, operating on NonMR+, NonMR- or any other well-defined way, is always the same, exactly as in the game above. In this rather realistic but transparent framework, logic is shown to be normative.

# Bibliography

[Fre1893] Gottlob Frege, *Grundgesetze der Arithmetik* (Paragraphen in moderner Notation), electronic version [korpora.org], Universität Duisburg-Essen, 2006.
[Ha2009] Hartry Field and Peter Milne, *The normative role of logic*, in *Proceedings of the Aristotelian Society*, sup. vol. LXXXIII, 2009, pp. 251-268.
[Leh1992] Lehmann and Magidor, *What does a conditional knowledge base entail?*, in arXiv:cs/0202022v1, previously published, with erratum, in *Journal of Artificial Intelligence*, Vol. 55 no.1 (May 1992), pp. 1-60.
[Mac2004] John MacFarlane, *In what sense (if any) is logic normative for thought?*, unpublished (last draft dated April 21, 2004; preprint presented at the 2004 Central Division

APA symposium on the normativity of logic.)

[Post1944] E. L. Post, *Recursively enumerable sets of positive integers and their decision problems*, in *Bulletin of the American Mathematical Society* 50 (1944), pp. 284-316.

[Rogers1967] Hartley Rogers, *Theory of recursive functions and effective computability*, McGraw-Hill, 1967.